\documentclass[aps,prl,twocolumn,tightenlines,superscriptaddress]{revtex4}
\usepackage{epsfig,rotating}
\usepackage{multirow}
\newcommand{\nuc}[2]{\hbox{$^{#1}$#2}}

\begin{document}
%\draft
\title{Nuclear Structure Towards $N=40$ \nuc{60}{Ca}: In-beam $\gamma$-ray
  Spectroscopy of \nuc{58,60}{Ti}} 

\author{A.\ Gade}
   \affiliation{National Superconducting Cyclotron Laboratory,
      Michigan State University, East Lansing, Michigan 48824, USA}
   \affiliation{Department of Physics and Astronomy,
      Michigan State University, East Lansing, Michigan 48824, USA}
\author{R.\ V.\ F.\ Janssens}
    \affiliation{Physics Division, Argonne National Laboratory,
      Argonne, Illinois 60439, USA}
\author{D.\ Weisshaar}
    \affiliation{National Superconducting Cyclotron Laboratory,
      Michigan State University, East Lansing, Michigan 48824, USA}
\author{B.\,A.\ Brown}
    \affiliation{National Superconducting Cyclotron Laboratory,
      Michigan State University, East Lansing, Michigan 48824, USA}
    \affiliation{Department of Physics and Astronomy,
      Michigan State University, East Lansing, Michigan 48824, USA}
\author{E.\ Lunderberg}
    \affiliation{National Superconducting Cyclotron Laboratory,
      Michigan State University, East Lansing, Michigan 48824, USA}
    \affiliation{Department of Physics and Astronomy,
      Michigan State University, East Lansing, Michigan 48824, USA}
\author{M.\ Albers}
     \affiliation{Physics Division, Argonne National Laboratory,
       Argonne, Illinois 60439, USA}
\author{V.\,M. Bader}
    \affiliation{National Superconducting Cyclotron Laboratory,
      Michigan State University, East Lansing, Michigan 48824, USA}
    \affiliation{Department of Physics and Astronomy,
      Michigan State University, East Lansing, Michigan 48824, USA}
\author{T.\ Baugher}
    \affiliation{National Superconducting Cyclotron Laboratory,
      Michigan State University, East Lansing, Michigan 48824, USA}
    \affiliation{Department of Physics and Astronomy,
      Michigan State University, East Lansing, Michigan 48824, USA}
\author{D.\ Bazin}
    \affiliation{National Superconducting Cyclotron Laboratory,
      Michigan State University, East Lansing, Michigan 48824, USA}
\author{J.\ S.\ Berryman}
    \affiliation{National Superconducting Cyclotron Laboratory,
      Michigan State University, East Lansing, Michigan 48824, USA}
\author{C.\,M.~Campbell}
      \affiliation{Nuclear Science Division, Lawrence Berkeley
          National Laboratory, California 94720, USA}
\author{M.\ P.\ Carpenter}
     \affiliation{Physics Division, Argonne National Laboratory,
       Argonne, Illinois 60439, USA}
\author{C.\ J.\ Chiara}
      \affiliation{Department of Chemistry and Biochemistry,
        University of Maryland, College Park, Maryland 20742, USA}
       \affiliation{Physics Division, Argonne National Laboratory,
         Argonne, Illinois 60439, USA}
\author{H.\ L.\ Crawford}
       \altaffiliation{Institute of Nuclear and Particle Physics, and Department
         of Physics and Astronomy, Ohio University, Athens, OH 45701, USA} 
        \affiliation{Nuclear Science Division, Lawrence Berkeley
          National Laboratory, California 94720, USA}
\author{M.\ Cromaz}
        \affiliation{Nuclear Science Division, Lawrence Berkeley
          National Laboratory, California 94720, USA}
\author{U.\ Garg}
    \affiliation{Department of Physics, University of Notre Dame,
      Notre Dame, Indiana 46556, USA}
\author{C.\ R.\ Hoffman}
        \affiliation{Physics Division, Argonne National Laboratory,
          Argonne, Illinois 60439, USA}
\author{F.\ G.\ Kondev}
       \affiliation{Nuclear Engineering Division, Argonne National
         Laboratory, Argonne, Illinois  60439, USA}
\author{C.\ Langer}
       \affiliation{National Superconducting Cyclotron Laboratory,
      Michigan State University, East Lansing, Michigan 48824, USA}
       \affiliation{Joint Institute for Nuclear Astrophysics, Michigan State
         University, East Lansing, MI 48824, USA} 
\author{T.\ Lauritsen}
       \affiliation{Physics Division, Argonne National Laboratory,
         Argonne, Illinois 60439, USA}
\author{I.\,Y.~Lee}
      \affiliation{Nuclear Science Division, Lawrence Berkeley
          National Laboratory, California 94720, USA}
\author{S.\ M.\ Lenzi}
      \affiliation{Dipartimento di Fisica e Astronomia dell'Universit\`{a}
  and INFN, Sezione di Padova, I-35131 Padova, Italy}
\author{J.\ T.\ Matta}
    \affiliation{Department of Physics, University of Notre Dame,
      Notre Dame, Indiana 46556, USA}
\author{F.\ Nowacki}
    \affiliation{IPHC, IN2P3-CNRS et Universit\`e de Strasbourg, F-67037
      Strasbourg, France} 
\author{F.\ Recchia}
    \altaffiliation{Dipartimento di Fisica e Astronomia ``Galileo
      Galilei'', Universit`a degli Studi di Padova, I-35131 Padova, Italy.}
    \affiliation{National Superconducting Cyclotron Laboratory,
      Michigan State University, East Lansing, Michigan 48824, USA}
\author{K.\ Sieja}
    \affiliation{IPHC, IN2P3-CNRS et Universit\`e de Strasbourg, F-67037
      Strasbourg, France} 
\author{S.\ R.\ Stroberg}
    \affiliation{National Superconducting Cyclotron Laboratory,
      Michigan State University, East Lansing, Michigan 48824, USA}
    \affiliation{Department of Physics and Astronomy,
      Michigan State University, East Lansing, Michigan 48824, USA}
\author{J.\ A.\ Tostevin}
     \affiliation{Department of Physics, Faculty of Engineering and Physical
       Sciences, University of Surrey, Guildford GU2 7XH, United Kingdom}
\author{S.\ J.\ Williams}
    \affiliation{National Superconducting Cyclotron Laboratory,
      Michigan State University, East Lansing, Michigan 48824, USA}
\author{K.\ Wimmer}
     \affiliation{Department of Physics, Central Michigan University,
       Mt. Pleasant, Michigan 48859, USA}   
     \affiliation{National Superconducting Cyclotron Laboratory,
      Michigan State University, East Lansing, Michigan 48824, USA}
\author{S.\ Zhu}
     \affiliation{Physics Division, Argonne National Laboratory,
       Argonne, Illinois 60439, USA}
\date{\today}

\begin{abstract}

Excited states in the neutron-rich $N=38,36$ nuclei \nuc{60}{Ti} and
\nuc{58}{Ti} were populated in nucleon-removal reactions from
\nuc{61}{V} projectiles at 90~MeV/nucleon. The $\gamma$-ray
transitions from such states in these Ti isotopes were detected
with the advanced $\gamma$-ray tracking array GRETINA and
were corrected event-by-event for large Doppler shifts ($v/c \sim 0.4$)
using the 
$\gamma$-ray interaction points deduced from 
online signal decomposition. The new data indicate that a steep decrease in
quadrupole collectivity occurs when moving from neutron-rich $N=36,38$ Fe and Cr
toward the Ti and Ca isotones. In fact, \nuc{58,60}{Ti} provide some of the most
neutron-rich benchmarks accessible today for calculations attempting to
determine the structure of the potentially doubly-magic nucleus
\nuc{60}{Ca}. 
\end{abstract}

\maketitle

One of the main goals of nuclear physics is the development of a predictive
model for the properties of all nuclei, including the shortest-lived
species 
in as yet unexplored regions of the nuclear chart. This is important, for
example, in the quest to understand the origin of the elements in the Universe
since many nucleosynthesis processes involve nuclei far
removed from the valley of $\beta$ stability.  One of the cornerstones in the
description of nuclear properties is nuclear shell structure --
whereby discrete nucleon single-particle orbitals are clustered in energy,
resulting in stabilizing energy gaps occurring
for certain ``magic'' proton or neutron numbers. Doubly-magic nuclei, with both
proton and neutron magic numbers, are particularly important for the
development of nuclear models as they serve as essentially inert cores, reducing
the many-body problem to that of the set of ``valence nucleons'' outside this
core. However, modifications of shell structure have already been 
observed in short-lived nuclei with extreme neutron-to-proton ratios, with
new shell gaps developing and some of the canonical magic numbers disappearing
\cite{Ots13,Sor08,Gad08r,Jan09}. Considerable experimental and theoretical
efforts 
are aimed at describing the physics driving such changes which are revealed
most clearly on the neutron-rich side of the nuclear chart.

Data for chains of proton-magic isotopes and regions of rapid
shell evolution offer (complementary) challenging tests of nuclear models,
allowing changes in nuclear structure to be tracked as a function of
isospin and providing demanding benchmarks for calculations incorporating new
physics effects. The chain of Ca isotopes (with magic proton number 
$Z=20$) and the region of neutron-rich nuclei near $N=40$, which are subject
to rapid shell and shape changes~\cite{Sor03,Cau04,Adr08,Gad10,Car13}, coincide
at \nuc{60}{Ca}. 
In addition to the first spin-orbit-driven neutron sub-shell 
closure at $N=28$ \nuc{48}{Ca}, the neutron-rich Ca isotopes exhibit two
additional sub-shell gaps at $N=32$~\cite{Pri01} and $N=34$~\cite{Jan02},
attributed in part to the action of the monopole parts of the proton-neutron
tensor force in the regime of large neutron excess~\cite{Ots01,Ots05}.

Nothing is known experimentally about the properties of the most neutron-rich
$N=40$ isotones 
\nuc{62}{Ti} and \nuc{60}{Ca}. While the existence of \nuc{62}{Ti} has been
established~\cite{Tar09}, \nuc{60}{Ca} has not yet been observed. In fact,
the position of the neutron drip line in Ca appears to depend sensitively on
both the  location 
of the neutron $1g_{9/2}$ orbital, which starts to be filled at $N=40$ in
\nuc{60}{Ca}, and on a variety of correlations and many-body
effects~\cite{Men01,Len10}. Calculations with realistic two- and three-body
forces~\cite{Hag12,Hol12} predict the neutron drip 
line to be located around \nuc{60}{Ca}, while many mean-field and
density-functional theories  
have the Ca isotopes (at least those with even $A$) bound out to $A=68-76$
\cite{Erl12}. The relativistic continuum Hartree-Bogoliubov approach of Meng
{\it et al.}~\cite{Men01} has the neutron $1g_{9/2}$ and $3s_{1/2}$ orbitals
unbound, but correlation effects, predominantly pairing, bind even-$A$
Ca out to $A=72$, while the SkM$^*$ Skyrme functional has the neutron
$1g_{9/2}$ orbital bound and predicts the Ca drip line to be at
$A=70$~\cite{Men01}. Clearly, information on the structure of neutron-rich
nuclei with $A \approx 60$ is important to help benchmark modern calculations
which differ in their prediction of the location of the Ca drip line by
more than 10 mass units. The  calculations of Ref.~\cite{Hag13} suggest that the
regime of weak binding applying to \nuc{60}{Ca} leads to intriguing consequences
such as the presence of a halo structure and of two Efimov states in
\nuc{62}{Ca}.   

The first spectroscopy of \nuc{60}{Ti} and the identification of
new $\gamma$-ray transitions in \nuc{58}{Ti} are reported
here. At present, $^{60}_{22}$Ti$_{38}$ is probably 
the nucleus closest to $^{62}_{22}$Ti$_{40}$
and $^{60}_{20}$Ca$_{40}$ that can be studied until next-generation rare-isotope
facilities come 
online. The measurements were enabled by the luminosity inherent to fast
fragmentation-beam measurements~\cite{Gad08r} and the efficiency and spectral
quality provided 
by the advanced $\gamma$-ray tracking array GRETINA~\cite{gretina}.

Excited states in the neutron-rich Ti isotopes were populated in the
\nuc{9}{Be}(\nuc{61}{V},\nuc{58,60}{Ti}$+\gamma$)X nucleon removal reactions
at 90.0~MeV/u at the Coupled Cyclotron Facility at NSCL. The \nuc{61}{V} ions
were 
produced from a 140-MeV/u primary \nuc{82}{Se} beam impinging on
a 423-mg/cm$^2$ \nuc{9}{Be} production target, and were separated using a
240-mg/cm$^2$ 
Al degrader in the A1900 fragment separator~\cite{a1900}. The
momentum acceptance of the separator was restricted to 3\%, yielding typical
on-target rates of 15 \nuc{61}{V}/s. About 10\% of the secondary beam was
\nuc{61}{V}, with \nuc{62}{Cr} (32\%) and \nuc{64}{Mn} (45\%) being the most
intense other components.

\begin{figure}[h]
        \epsfxsize 8.0cm
        \epsfbox{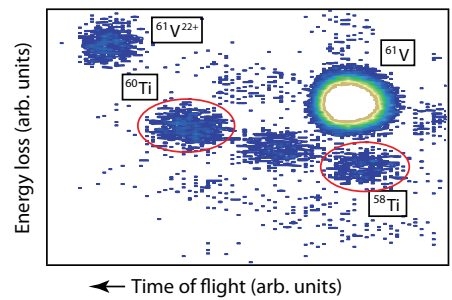}
\caption{\label{fig:pid} (Color online) Identification spectrum for
  the reaction residues produced in \nuc{9}{Be}(\nuc{61}{V},\nuc{A}{Ti})X at
  a 90-MeV/u mid-target energy. All reaction residues are
  unambiguously identified by their energy loss, measured in the
  S800 ionization chamber, and their time of flight.}
\end{figure}

The secondary \nuc{9}{Be} reaction target (376~mg/cm$^2$ thick) was located at
the target position of the S800 spectrograph~\cite{s800}. Reaction products were
identified 
on an event-by-event basis at the instrument's focal plane with the standard
detection system~\cite{s800}. The particle-identification spectrum for
\nuc{58,60}{Ti} produced from incoming \nuc{61}{V} ions is presented in
Fig.~\ref{fig:pid}. The spectrograph 
was centered on the \nuc{60}{Ti} one-proton knockout residues while the
\nuc{58}{Ti} momentum distribution was cut by the S800 acceptance. The inclusive
cross section for one-proton knockout from \nuc{61}{V} to \nuc{60}{Ti} was
measured to be $\sigma_{inc}=7.9(7)$~mb.

\begin{figure}[h]
        \epsfxsize 8.2cm
        \epsfbox{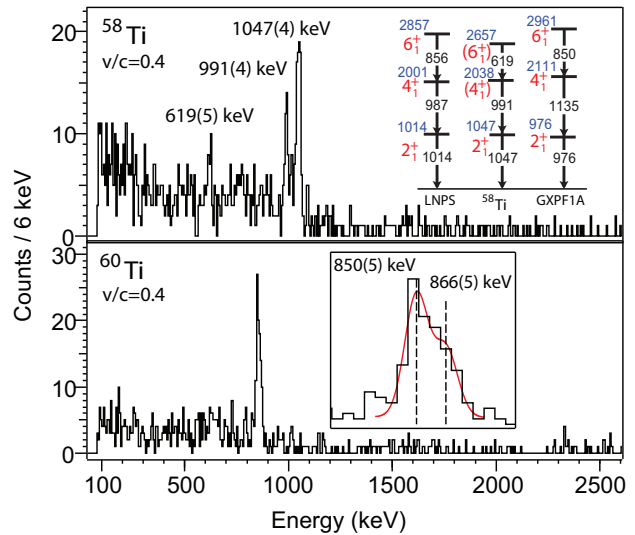}
\caption{\label{fig:gamma} Doppler-reconstructed $\gamma$-ray
  spectra in coincidence with \nuc{58,60}{Ti} reaction residues. The indication
  of a transition doublet in \nuc{60}{Ti} is shown as an inset in the lower
  panel.} 
\end{figure}

The high-resolution $\gamma$-ray detection system
GRETINA~\cite{gretina}, an array of 36-fold segmented high-purity Ge detectors,
was used to measure the prompt $\gamma$ rays emitted by the reaction
residues. The seven GRETINA  modules -- with four crystals each -- were
arranged in two rings. Four modules were located at
58$^{\circ}$ and three at 90$^{\circ}$ with respect to the beam axis. Online
signal decomposition provided $\gamma$-ray interaction points for event-by-event
Doppler reconstruction of the photons emitted in-flight at $v/c= 0.4$. The
information on the momentum vector of projectile-like reaction residues, as
reconstructed from ray-tracing through the spectrograph, was
incorporated in the Doppler reconstruction. Figure~\ref{fig:gamma} presents
these Doppler-reconstructed spectra for \nuc{58}{Ti} and
\nuc{60}{Ti} with addback included~\cite{Wei13}. The high 
peak-to-background ratio enables spectroscopy to be performed at the low levels
of statistics that are inherent to investigations of the most exotic nuclei. 

In \nuc{58}{Ti}, in addition to the previously known $2^+_1 \rightarrow 0^+_1$
transition at 1047(4)~keV~\cite{Aoi08,Suz13}, two additional $\gamma$ rays could
be 
identified 
at 991(4)~keV and 619(5)~keV. The reaction to \nuc{58}{Ti} is not a direct
process and, typically, the population of yrast states is favored in such 
fragmentation processes. Thus, based on the observed intensity pattern, the
991-keV $\gamma$ ray is tentatively assigned as the $4^+_1 \rightarrow 2^+_1$
decay and it 
is suggested  that the 619-keV line corresponds to the $6^+_1 \rightarrow 4^+_1$
transition. Evidence for the 
coincidence between the 991- and 1047-keV $\gamma$ rays is discussed below. The
inset shows that the two shell-model effective interactions for this region of
the nuclear chart, LNPS~\cite{Len10} and GXPF1A~\cite{Honma}, describe the
proposed level scheme well 
although the neutron model spaces differ significantly with GXPF1A restricted to
the neutron $fp$ shell and LNPS including the neutron $d_{5/2}$ and $g_{9/2}$
orbitals in addition.  The agreement is good with both interactions, suggesting
that incorporating the $1g_{9/2}$ and $2d_{5/2}$ neutron orbitals may not be
critical, in agreement with conclusions presented in~\cite{Suz13}. Note, that
the 
present spin assignments 
are consistent with the fact that the 991- and 619-keV transitions were not
observed by \cite{Suz13}, since $4^+$ and $6^+$ states are not expected to be
populated strongly in inelastic proton scattering.

In \nuc{60}{Ti}, a peak structure at 860 keV is observed on top of very little
background. One-proton knockout is a direct reaction with sensitivity to the
single-particle degrees of freedom and it offers insight into the overlap in 
structure between the projectile ground state and the final states populated in
the knockout residue~\cite{knock}. The partial cross section to an excited final
state is 
determined from the efficiency-corrected peak area relative to the
number of  
knockout residues. A GEANT4 simulation of the GRETINA setup~\cite{geant}, that
reproduced the intensity of standard calibration sources, was used to model the
in-beam full-energy peak efficiency of the detector array, including the
Lorentz boost. The simulated in-beam efficiency was employed to extract
intensities from the peak areas in \nuc{60}{Ti}. Assuming that the peak
structure in 
\nuc{60}{Ti} corresponds to a single transition then implies that 111(12)\% of
the 
knockout proceeds to the state depopulated by this 860-keV transition and that
there is essentially no  population of any other final state in
\nuc{60}{Ti}. For a 
nucleus bound by more than 5~MeV, this scenario appears to be rather unlikely.

\begin{figure}[h]
        \epsfxsize 8.5cm
        \epsfbox{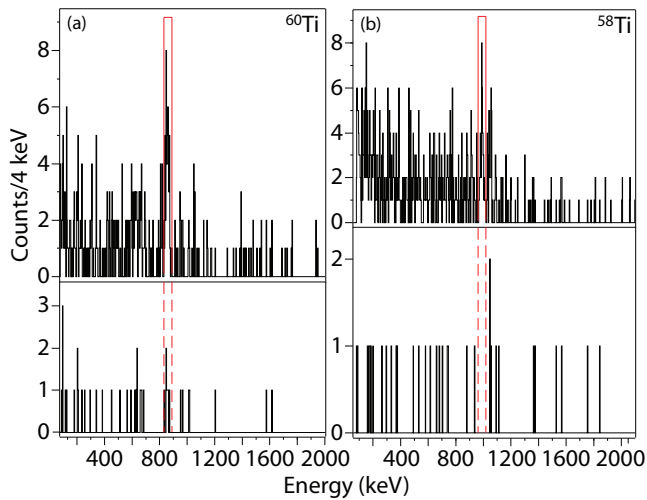}
\caption{\label{fig:coinc} (Color online) Projections of the \nuc{60,58}{Ti}
  $\gamma\gamma$ coincidence matrices, nearest-neighbor addback
  included~\cite{Wei13} (upper
  panels) and gated coincidence spectra (lower panels): (a) \nuc{60}{Ti} -- the
  gate on the 860-keV peak returns a self-coincidence and Compton
  edge; (b) \nuc{58}{Ti} -- the gate on the 991-keV transition shows no
  self-coincidence and returns the peak at 1047~keV, consistent with a
  1047-991-keV cascade.}
\end{figure}

In fact, the asymmetric peak shape at 860~keV supports the presence of a doublet
(Fig.~\ref{fig:gamma}, inset). Analysis as a doublet suggests the presence of
two $\gamma$ rays at  
850(5)~keV  and 866(5)~keV, presumably corresponding to the $2^+_1 \rightarrow
0^+_1$ and $4^+_1 \rightarrow 2^+_1$ transitions in
\nuc{60}{Ti}, associated with 40(10)\% population of the $4^+_1$ state, 30(11)\%
of the $2^+_1$ level and 29(12)\% of the ground state and unobserved levels not
feeding the proposed $2^+_1$ or $4^+_1$ states. GRETINA's $\gamma\gamma$
coincidence capability supports further examination 
of the proposed doublet. Figure~\ref{fig:coinc}(a) provides the total projection
of the coincidence matrix for \nuc{60}{Ti} (upper panel) and the spectrum gated
on the 860-keV peak (lower panel). Clearly, a peak in the same region
(self-coincidence) and a corresponding Compton edge between 600-700~keV are
visible. Similarly, in Fig.~\ref{fig:coinc}(b), the projection of the
\nuc{58}{Ti} $\gamma\gamma$ coincidence matrix is given (upper panel) as is the
spectrum with a coincidence condition on the 991-keV transition (lower
panel). No self-coincidence events are 
visible; instead the 1047-keV $\gamma$ ray appears, confirming the 991-1047-keV
cascade. Thus, the self-coincidence of the 
860-keV peak structure is evidence for a coincident doublet of $\gamma$-ray
transitions in \nuc{60}{Ti}. 

\begin{figure}[h]
        \epsfxsize 8.6cm
        \epsfbox{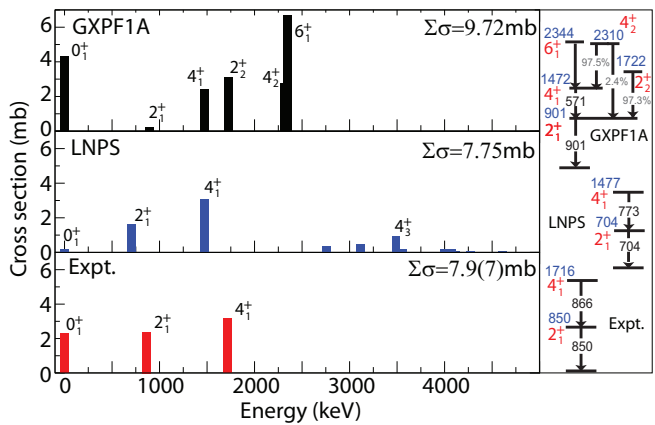}
\caption{\label{fig:xsec} (Color online) Calculated and measured partial cross
  sections to final states in \nuc{60}{Ti} using GXPF1A and LNPS spectroscopic
  factors and the procedure outlined in the text. The experimental cross
  sections 
  to the $2^+_1$ and $4^+_1$ states were deduced from the $\gamma$-ray
  intensities and the $0^+_1$ population results from subtraction. The so
  determined strength constitutes 29(12)\%, 30(11)\% and 40(10)\% population of
  the $0^+_1$, $2^+_1$ and $4^+_1$ states, which will include unobserved
  feeding from higher-lying states that could not be observed due to the lack of
statistics (In Fig.~\ref{fig:gamma}, about 8 counts should be seen in the
\nuc{60}{Ti} spectrum per 1~mb at 1.5 MeV).}    

\end{figure}

Knockout calculations can be used for further guidance. The ground-state spin of
\nuc{61}{V} is not known experimentally. Shell-model calculations  
with the LNPS effective interaction predict a $3/2^-$ ground state, in
agreement with $\beta$-decay results~\cite{Gau05}. The GXPF1A effective
interaction~\cite{Honma}, 
which does not include the potentially important neutron $d_{5/2}$ and $g_{9/2}$
orbitals, 
predicts a $7/2^-$
ground state with excited $5/2^-_1$ and $3/2^-_1$ levels within
400~keV.  Using the  
one-nucleon knockout formalism detailed in~\cite{Gad08},  the
GXPF1A and LNPS spectroscopic factors with respect to the ground
state of \nuc{61}{V}, and assuming a reduction factor of $R_s \approx 0.5$ at a
nucleon separation 
 energy difference of the projectile of $S_n-S_p \approx -10$~MeV~\cite{Gad08},
 the partial cross sections to bound final states in \nuc{60}{Ti} are calculated
 and confronted with experiment in Fig.~\ref{fig:xsec}. For the LNPS effective
 interaction, the calculated inclusive cross section agrees with the
 measurement, while the GXPF1A calculation predicts a slightly higher cross
 section. From the GXPF1A calculation, four excited levels, $4^+_1$, $2^+_2$,
 $4^+_2$, and $6^+_1$, should be populated strongly. There is 
 no  evidence in the spectrum for additional strong $\gamma$ rays that would
 correspond to the respective transitions. Note that assuming a $5/2^-$ or
 $3/2^-$ ground state within the GXPF1A calculations always results in the strong
 population  of three or more excited states, corresponding to the presence of
 several strong $\gamma$-ray transitions in addition to the $2^+_1 \rightarrow
 0^+_1$ and $4^+_1 \rightarrow 2^+_1$ decays (e.g. the $6^+_1 \rightarrow
 4^+_1$, $2^+_2 \rightarrow 2^+_1$, or the $4^+_2 \rightarrow 4^+_1$
 transitions). With LNPS, the
 single-particle strength distribution resembles the data with the majority of
 the cross section carried by the $2^+_1$ and $4^+_1$ states. Discrepant is the
 $29(12)$\% population deduced for the ground state by subtraction. However, the
 experimental strength to the ground state will also include unobserved feeding
 from higher excited levels that bypass the $2^+_1$ and $4^+_1$ states and
 will act as a funnel for a fraction of the strength predicted to be
 fragmented over higher-lying states.  Unlike for
 the \nuc{60}{Ti} excitation energies, which do not signal a clear difference
 between the 
 predictions for the different model spaces, the spectroscopic strengths clearly
 indicate that the neutron $d_{5/2}$ and $g_{9/2}$ orbitals are important for
 the description of the overlap of the \nuc{61}{V} ground state with the
 final-state wave functions in \nuc{60}{Ti}.

Over the past decade, with advances in nuclear experiment, the
neutron-rich Cr and Fe nuclei around $N=40$
were found to be strongly deformed, presenting a challenging testing ground
for the theoretical modeling of shell evolution. The next experimental and
theoretical key benchmark is the understanding of the neighboring $N=2Z$ nucleus
\nuc{60}{Ca}. As shown in Fig.~1 of Ref.~\cite{Len10}, 
the $N=40$ shell gap computed with the LNPS interaction vanishes as $Z=20$
(calcium) is 
approached, and the ground state of $^{60}$Ca is dominated by
four-particle four-hole 
($4p4h$) neutron excitations from the $pf$ shell into the $g_{9/2}$-$d_{5/2}$
orbitals (Table I in \cite{Len10}). However, the shell-model extrapolation of
single-particle  energies is often not 
accurate~\cite{Bro01}, perhaps due to the lack of inclusion of three-body
forces. Another approach would be to use Hartree-Fock or
Energy-Density-Functional (EDF) 
calculations to estimate the $N=40$ sub-shell gap. The 12 CSkP Skyrme
functionals, used in~\cite{Bro13}, give a shell gap between the neutron
$f_{5/2}$ and $g_{9/2}$ orbitals varying between 3 and 4~MeV at $Z=20$. If
the shell gap were this large, the ground state of $^{60}$Ca would be dominated
by $0p0h$ rather than $4p4h$ configurations. Clearly, the size
of the $N=40$ shell gap is crucial for the properties of nuclei in
this region. Collective nuclei, such as $^{64}$Cr, are in the ``island of
inversion''~\cite{Len10} because of strong quadrupole correlations
for {\it both} protons and neutrons. With a large shell gap, there would be a
dramatic change from a deformed to a spherical shape as one approaches $Z=20$,
since the protons encounter a spin-orbit (LS) closed shell with no available
low-lying proton quadrupole excitations.

Shell-model calculations with the LNPS interaction provide a good description of
the data in this region. In the case of \nuc{60}{Ti}, the
excitation energies of both states are underestimated ($2^+_1$ energy by 150~keV
and the $4^+_1$ one by 240~keV). Since this nucleus is one of the furthest
extrapolation points with no data available previously, it is interesting
to study its sensitivity to modifications of the interaction and the resulting
impact on the calculated structure of this region. Such modifications to the
LNPS 
effective interaction -- based on available, independent data in this region --
are 
underway and offer the opportunity to assess the role of
\nuc{60}{Ti}. With an increase of the $d_{5/2}$ single-particle energy by 250 keV
and repulsion of $g_{9/2},d_{5/2}$ monopole matrix elements by 200 keV, the description of the
excitation energies of \nuc{60}{Ti} improves, with the $2^+_1$ state calculated at 803 and the $4_1^+$ level at
1609~keV. While these modifications increase the small $N=40$ gap at
\nuc{60}{Ca} by 
only 250~keV, they significantly alter the nuclear structure of the
region with markedly changed $2p2h$ and $4p4h$ contributions to the wave
functions. In the original LNPS effective interaction, the ground state and
$2^+_1$ state of \nuc{60}{Ti} contain 27\% of $2p2h$ and 41\% of $4p4h$ and
15\% of $2p2h$ and 
45\% $4p4h$ contributions, respectively. With the modifications that improve the
agreement for the \nuc{60}{Ti} excitation energies, these contributions
change to 36\% of $2p2h$ and 33\% of $4p4h$ and 21\% of $2p2h$, 39\% of $4p4h$
for the ground and $2^+_1$ states, respectively. Confirmation of the size of
the $N=40$ shell gap and of the role of multi-particle multi-hole configurations
beyond \nuc{60}{Ti} will likely only come with the next generation of
experiments measuring properties of nuclei even closer to \nuc{60}{Ca} combined
with advances in nuclear theory such as improved effective shell-model
interactions built on those developed currently.

In summary, first structural information on \nuc{60}{Ti} was obtained by taking
advantage of the spectral quality and the $\gamma$-ray coincidence efficiency of
GRETINA. The first $2^+_1$ state of \nuc{60}{Ti}, at an energy of 850(5)~keV, is
located at almost twice the excitation energy of the corresponding $2^+_1$ level
in the $N=38$ isotone \nuc{62}{Cr}, herewith signaling a steep decrease in
collectivity with $Z$ and yet another sudden structural change near $N=40$. For 
\nuc{58}{Ti}, candidates for the $(4^+_1)$ and $(6^+_1)$ levels are reported. The
data on \nuc{60}{Ti} are consistent with a shell-model prediction using the LNPS
effective interaction which allows for the largest neutron model space yet,
while they disagree with calculations restricted to the 
neutron $fp$ shell. The \nuc{60}{Ti} excitation energies were shown to be
sensitive to the details of the effective interaction, with significant impact
on the particle-hole contents of the state's wave functions. This in turn drives
the nuclear structure in this region. With this, \nuc{60}{Ti} represents 
an important benchmark, being one of the most neutron-rich systems from
which to extrapolate towards \nuc{60}{Ca}, a nucleus with an intrinsic structure
closely tied to the location of the neutron drip line in the crucial semi-magic
Ca isotopic chain. 

\begin{acknowledgments}
GRETINA was funded by the DOE, Office of Science. Operation of the array at
NSCL was supported by NSF under Cooperative Agreement PHY-1102511 (NSCL) and DOE
under Grant DE-AC02-05CH11231 (LBNL). We further acknowledge DOE Contract 
DE-AC02-06CH11357 (ANL) and Grant DE-FG02-94-ER40834 (UM) and support from NSF
grant PHY-1068217 (NSCL) and PHY-1068192 (ND). J.A.T. acknowledges
support of the Science and Technology Facilities Council (UK) grant
ST/J000051. We are grateful for Augusto Macchiavelli's support of the GRETINA
campaign at NSCL.
\end{acknowledgments}

\end{document}